\documentclass[journal=jpccck,manuscript=article]{achemso}

\usepackage[version=3]{mhchem} 
\usepackage{lineno,xcolor}

\author{Jessica K. Bristow}
\affiliation{Centre for Sustainable Chemical Technologies and Department of Chemistry, University of Bath, Claverton Down, Bath BA2 7AY, UK}
\author{Katrine L. Svane}
\affiliation{Centre for Sustainable Chemical Technologies and Department of Chemistry, University of Bath, Claverton Down, Bath BA2 7AY, UK}
\author{Davide Tiana}
\affiliation{Centre for Sustainable Chemical Technologies and Department of Chemistry, University of Bath, Claverton Down, Bath BA2 7AY, UK}
\author{Jonathan M. Skelton}
\affiliation{Centre for Sustainable Chemical Technologies and Department of Chemistry, University of Bath, Claverton Down, Bath BA2 7AY, UK}
\author{Julian D. Gale}
\affiliation{Nanochemistry Research Institute/Curtin Institute for Computation, Department of Chemistry, Curtin University, PO Box U1987, Perth, WA 6845, Australia}
\email{j.gale@curtin.edu.au}
\author{Aron Walsh}
\affiliation{Centre for Sustainable Chemical Technologies and Department of Chemistry, University of Bath, Claverton Down, Bath BA2 7AY, UK}
\affiliation{Global E$^3$ Institute and Department of Materials Science and Engineering, Yonsei University, Seoul 120-749, Korea}
\email{a.walsh@bath.ac.uk}
\title[\texttt{achemso} demonstration]
{Free energy of ligand removal in the metal-organic framework UiO-66}

\begin{document}

\begin{abstract}
We report an investigation of the ``missing-linker phenomenon'' in the Zr-based metal-organic framework UiO-66 using atomistic forcefield and quantum chemical methods. For a vacant benzene dicarboxylate ligand, the lowest energy charge capping mechanism involves acetic acid or Cl$^{-}$/H$_{2}$O. The calculated defect free energy of formation is remarkably low, consistent with the high defect concentrations reported experimentally. A dynamic structural instability is identified for certain higher defect concentrations. In addition to the changes in material properties upon defect formation, we assess the formation of molecular aggregates, which provide an additional driving force for ligand loss. These results are expected to be of relevance to a wide range of metal-organic frameworks. 
\end{abstract}

\section{Introduction}

Metal-organic frameworks (MOFs) are materials formed \textit{via} the coordination of metal centres and organic linkers in three dimensions. 
The varied chemical compositions and structural topologies of MOFs make them suitable for a broad range of applications including gas storage and separation, solar energy conversion, and heterogeneous catalysis.\cite{ma2009enantioselective, liu2015photoinduced, arstad2008amine, alvaro2007semiconductor, hwang2008amine, liu2007assembly}

\begin{figure}
\centering
\includegraphics[width=8.3cm]{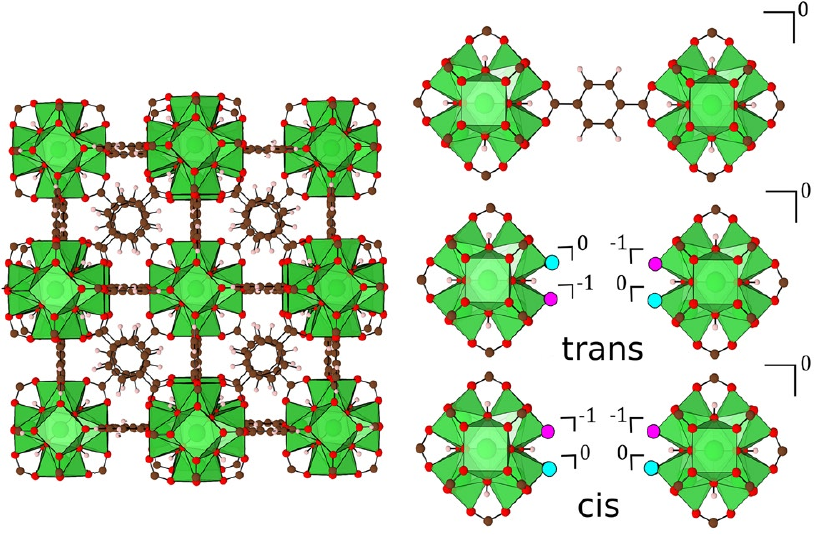}
\caption{The crystal structure of UiO-66 (left) and locations on the metal node where charge compensating or neutral molecules can bind following BDC linker removal (right). The locations of charge compensating molecules are highlighted in maroon and neutral molecule locations are highlighted in black. Top right shows the BDC linker connection between Zr-metal nodes prior to removal. Centre and bottom right shows the locations considered for charge compensating molecules following linker removal.}
\label{3mofs}
\end{figure}

One MOF that has attracted particular attention is UiO-66 (Figure \ref{3mofs}), which was first synthesised by Cavka \textit{et al.} \cite{cavka2008new}. 
This material features a high coordination of 12 benzene-1,4-dicarboxylate (BDC) ligands around each Zr$^{IV}$ node and is thermally stable up to 813 K.\cite{valenzano2011disclosing} 
The internal surface area (800 m$^{2}$g$^{-1}$) is large, with the structure containing both tetrahedral and octahedral cages. 
Each octahedral cage is edge-sharing with 8 tetrahedral cages and face-sharing with 8 octahedral cages.\cite{katz2013facile} 
The inner-sphere coordination of Zr in UiO-66 is 6, but additional face-sharing oxide and hydroxide ligands lead to an outer-sphere coordination of 12. 
%

Wu \textit{et al.} and Vermoortele \textit{et al.} reported a significant internal surface area increase for UiO-66 synthesised with an acidic modulator such as acetic or hydrochloric acid.\cite{vermoortele2013synthesis, wu2013unusual} 
This phenomenon, leading to increased gas storage capabilities with little stability loss, has been attributed to a missing BDC linker from the unit-cell, with a subsequent reduction in coordination of the Zr metal.\cite{van2015improving, barin2014defect, katz2013facile} 
The acid modulator has been shown to promote linker removal.\cite{vermoortele2013synthesis} 
Recent reports have focused on the charge capping mechanism following the removal of the linker. 
Experimental evidence, such as quantum tunnelling peaks in inelastic neutron scattering, associated with terminating methyl groups, suggest acetic acid becomes incorporated into the framework.\cite{wu2013unusual} 
The incorporation of Cl$^{-}$ ions when using HCl has also been suggested.\cite{shearer2014tuned} 
Considering that an excess of ZrCl$_{4}$ is often used during synthesis and that experimental conditions do not completely exclude water, there is an abundance of potential charge capping ions. 

NU-1000\cite{planas2014defining} is a structurally similar Zr-containing MOF, which is often compared to UiO-66. The Zr node in NU-1000 has the formula [Zr$_{6}$($\eta_{3}$-O)$_{4}$($\eta_{3}$- OH)$_{4}$(OH)$_{4}$(H$_{2}$O)$_{4}$]$^{8+}$, and in UiO-66 [Zr$_{6}$($\eta_{3}$-O)$_{4}$($\eta_{3}$-OH)$_{4}$]$^{12+}$. The additional incorporation of four hydroxide and four water molecules in NU-1000 is due to the use of ZrOCl$_{2}$ as the Zr precursor source, as opposed to the ZrCl$_{4}$ precursor used to synthesise UiO-66.\cite{hod2014directed, yang2015metal, planas2014defining} Indeed, NU-1000 is an example of an ordered defect structure. 

The fraction of BDC linkers missing from UiO-66 is highly debated. Reports vary from 1--4 vacancies per metal node depending on synthesis conditions; however, all measurements are indirect (e.g. thermogravimetric analysis) and usually yield an average over a large sample volume.
 Regardless of the method employed, it is clear that the defect concentrations are high and beyond those typically found in crystalline materials.

In this paper, we investigate the free energy of formation of missing ligand defects in UiO-66 using a combination of first-principles and molecular mechanics computational techniques.
 We consider a range of charge compensating schemes involving commonly used species. 
The results validate recent experimental observations of high defect concentrations and reveal a thermodynamic driving force for defect aggregation in the UiO-66 system. 

\section{Methodology}

The predictive power of computational chemistry applied to metal-organic frameworks is well established. 
Here, we combine empirical and first-principle methods. 
The analytical forcefield calculations allow us to probe large and complex defect structures including vibrations, and hence calculate the Gibbs free energy of ligand removal.
The higher-level density functional theory calculations provide a means of validation, while also giving an estimate of solvation and cluster energies for reaction products that are challenging to compute using empirical interatomic potentials.

\subsection{Forcefield calculations}
We have considered the cubic unit cell of UiO-66, which contains 24 linkers and 4 metal nodes. 
Forcefield calculations were performed with \textsc{GULP}\cite{gale2003general, gale1997gulp}. 
Parametrisation of the interatomic potential was conducted to recreate the structural and material properties of non-defective UiO-66, including bond lengths, bond angles, phonon frequencies, bulk modulus and elastic constants. 
The details of the forcefield and a comparison of the predicted structure of UiO-66 against experimental data is given in the Supporting Information (SI). 
The bulk and defective structures were first optimized with respect to the internal energy, 
and then the free energy of the final structure was calculated including the vibrational entropy. 
For all defect reactions considered, reactants and products were optimised at constant external pressure, thus providing the Gibbs free energy ($\Delta$G) of reaction.

\subsection{Density functional theory calculations}
Reference solid-state density functional theory (DFT) calculations on the pristine and defective structures of UiO-66 were performed using \textsc{VASP}\cite{kresse1993ab}. 
These periodic DFT calculations were to provide high-quality fitting data for the forcefield and to validate the defect structures.  
The PBEsol functional \cite{perdew1996generalized} was used with a plane-wave cutoff of 600 eV and wavefunctions were calculated at the $\Gamma$ point of the Brillouin zone. 
Projector augmented wave potentials were used to model the interaction between valence and core of all atoms, with 5$s$4$d$5$p$
as the valence configuration of Zr.
Internal forces were converged to less than 0.005 eV/$\mathrm{\AA}$. 
The optimized unit-cell parameters from PBEsol/DFT ($a$ = 20.80 $\mathrm{\AA}$ and $\alpha = $90.0$^{\circ}$) reproduce the experimental structure ($a$ = 20.98 $\mathrm{\AA}$ and $\alpha = $90.0$^{\circ}$) of UiO-66 to within 1$\%$. 
Comparisons of the crystal structures produced by DFT and forcefield methods are given in the SI.

Free energies of solvation for molecular fragments in DMF (dimethylformamide) were calculated with the continuum solvation model, COSMO, in \textsc{NWChem}\cite{valiev2010nwchem} (cc-pVTZ basis set).\cite{dunning1989gaussian,woon1993gaussian} 
The self-consistent field energy convergence was set to 10$^{-6}$ Ha and the M06-2X functional\cite{zhao2008m06,zhao2008density}, which is known to produce accurate thermodynamic properties, was used to obtain geometries. 
In the solvation model we used the temperature-dependent experimental dielectric constant of DMF, as reported by Bass \textit{et al.}.\cite{bass1964dielectric}
Other thermodynamic quantities, such as the energy of protonation of BDC, were taken from the NIST database.\cite{lemmon2003nist}
Finally, molecular cluster binding energies were calculated with the B3LYP functional\cite{becke1993density}. This approach gives a good description of hydrogen bonding interactions at low computational cost. The dielectric constant of DMF at 300 K was used. A single point counterpoise correction for the basis set superposition error (BSSE)\cite{xantheas1996importance} was calculated on the converged cluster geometries.   

\section{Results}

\subsection{Charge capping mechanism}
For a balanced defect reaction, conservation of charge and mass is required.
Acetic acid (CH$_{3}$COOH) and/or HCl are commonly used as acidic modulators to promote linker removal from the structure. 
In addition, the commonly used solvent, DMF, and also H$_{2}$O can be incorporated. 
The removal of one BDC linker results in a system with an overall +2 charge, and reduces the coordination sphere of 4 Zr centres from 12 to 11. 
We consider seven capping mechanisms for charge compensation and stabilising the structure by saturating the coordination of each metal centre with a neutral molecule (Table \ref{chgmodels}). 

There are two choices for adding the charge capping and neutral molecules into the structure, labelled as trans and cis in Figure \ref{3mofs}. 
We find the lowest energy arrangement for trans substitution, which can be understood from simple electrostatics, as it maximises the distance between the charge capping species, and also steric effects. 
All results refer to the most stable (trans) configuration.

\begin{table}[h]
\caption{Charge compensation models for a missing linker from UiO-66. Given are the charge compensating molecules coordinated onto the two Zr centres, the precursors, and the neutral molecules included in some models to saturate the Zr coordination spheres.}
\centering
\begin{tabular}{c c c c} 
\hline
 Model & \multicolumn{2}{c}{Charged} & Neutral \\
 & Precursor & Anion &\\
\hline
1 & HCl & Cl$^{-}$ &  - \\
2 & HCl & Cl$^{-}$ & H$_{2}$O\\
3 & CH$_{3}$COOH & CH$_{3}$COO$^{-}$ & - \\
4 & H$_{2}$O & OH$^{-}$ &  -\\
5 & H$_{2}$O & OH$^{-}$ &  H$_{2}$O\\
6 & H$_{2}$O & OH$^{-}$ &  DMF\\
7 & HCl & Cl$^{-}$ &  DMF\\
\hline
\end{tabular}
\label{chgmodels}
\end{table}

\subsection{Defect formation energies}

\begin{figure*}
\centering
\includegraphics[width=12cm]{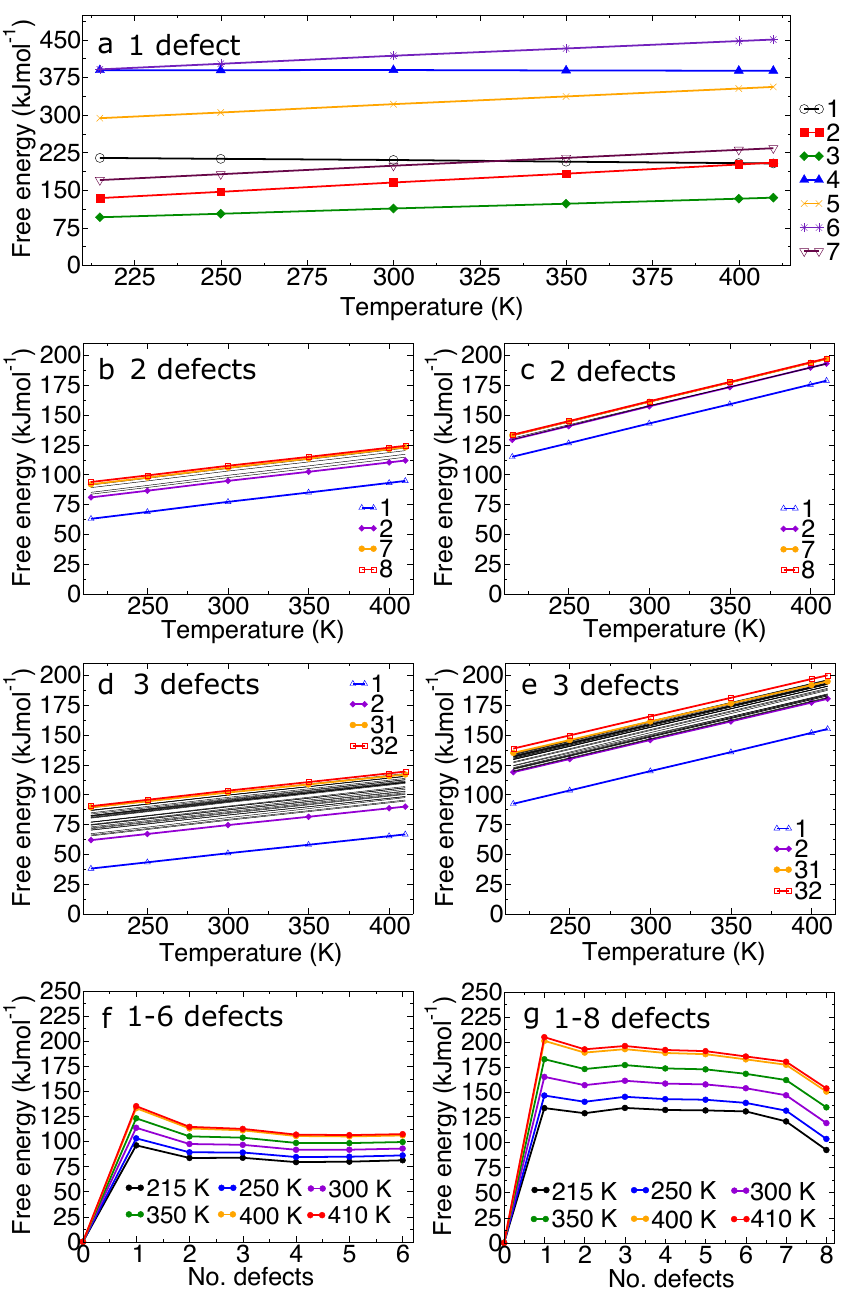}
\caption{Free energy of defect formation for:
(a) a single vacancy with a range of capping models (labelled 1--7 corresponding to Table \ref{chgmodels}).
(b -- c) two vacancies with acetate and Cl$^{-}$/H$_{2}$O capping models. Highlighted are the two lowest and highest energy configurations, all other configurations are shown as black lines.
(d -- e) three vacancies with acetate and Cl$^{-}$/H$_{2}$O capping models. 
(f -- g) removal of 1 -- 8 ligands for the CH$_{3}$COO$^{-}$ and Cl$^{-}$/H$_{2}$O charge capping models towards the formation of the ordered vacancy NU-1000 configuration.
All energies are presented per defect and include contributions from the vibrational entropy and zero point contribution to the enthalpy.}
\label{figure3}
\end{figure*}

The defect free energies as a function of temperature, calculated using mass and charged balanced chemical reactions, are given in Figure \ref{figure3}. 
The charge compensating models are detailed in Table \ref{chgmodels} and full reactions are listed in the SI. 
The reaction energy is sensitive to the charge compensation model. 
The inclusion of OH$^{-}$ as a binding ligand is particularly unfavourable. 
The higher calculated defect energy associated with OH$^{-}$ is due to the energy required to split its precursor (water) in DMF as a solvent. 

The charge capping mechanisms that had the lowest associated formation free energy were with acetic acid and Cl$^{-}$/H$_{2}$O. 
The acetic acid cap was optimised from multiple initial configurations. 
In each case the CH$_{3}$COO$^{-}$ ligand converged to a structure with bidentate coordination and identical bond lengths. Little structural distortion or loss of symmetry occurs to the framework of UiO-66 with the incorporation of acetic acid, due it possessing an identical head group to BDC. Slight losses of symmetry calculated when using CH$_{3}$COO$^{-}$ as the charge capping ion, are due to the loss of a mirror plane from the introduction of the methyl group. It is therefore the similarity between the chemical structure and solvation energies of the BDC and acetate head groups that makes acetic acid the lowest energy charge capping mechanism in UiO-66. 

Interestingly, we found that binding a Cl$^{-}$ ion with a neutral molecule had a much lower energy than binding only Cl$^{-}$ ions. Following the insertion of a monodentate charge capping ion alone we observed it bridging between two neighbouring Zr centres. When water/DMF were introduced, such that the Zr centres remained fully coordinated, the defect energy was lowered. This confirms, as expected, that an undercoordinated metal centre is energetically unfavourable. Our findings also suggest that a small concentration of water during synthesis may increase the number of linker vacancies within the material. We found the effect of coordinating DMF as a neutral molecule to have little influence on the defect energy. It can be seen that when comparing the energies for single Cl$^{-}$ and Cl$^{-}$/DMF substitution, DMF, as a neutral coordinating molecule, lowers the defect energy of removing one BDC linker. Note that between 350--400 K the energies of the respective charge capping mechanisms cross and the single Cl$^{-}$ model becomes more favourable than the Cl$^{-}$/DMF model; suggesting DMF coordination to be unfavourable at high temperatures. 

\subsection{Multiple ligand vacancies}

Taking the lowest energy charge capping mechanisms (CH$_{3}$COO$^{-}$ and Cl$^{-}$/H$_{2}$O), as identified in Figure \ref{figure3}(a), we further investigated the defect energies associated with the removal of additional BDC ligands. We present the defect energies for each of the symmetry unique locations of 2 BDC removals in Figure \ref{figure3}(b--c). The details of these configurations are given in the SI. 
The lowest energy configurations are identified to occur when removing linkers from the faces of the same tetrahedral cage, which also form the vertices of the central octahedral cage. The most favourable position renders one metal node as 10 coordinate and two other metal nodes as 11 coordinate. 

For the removal of three BDC linkers, we calculate 32 symmetry unique configurations in a single unit cell. We have calculated the defect formation energy of all configurations for the lowest energy charge capping mechanisms (CH$_{3}$COO$^{-}$ and Cl$^{-}$/H$_{2}$O), Figure \ref{figure3}(d--e). Each configuration is numbered in order of increasing magnitude of the defect energy, (\textit{i.e.} configuration 1 has the lowest energy and configuration 32 has the highest). We find a broader distribution of defect energies for the acetate capping than for Cl$^{-}$/H$_{2}$O. 
We observe the short-range structural disorder in the acetate configurations, where the acetate molecule pints into the pore and does not stay in planar alignment, to be larger with clustered defects due to local interactions and a loss of symmetry. Configuration 1 has the lowest defect energy by 23.8 and 26.5 kJmol$^{-1}$ for the acetate and Cl$^{-}$/H$_{2}$O capping, respectively, when compared to configuration 2. This configuration corresponds to three BDC linkers being removed from the same tetrahedral cage within the structure, with strong local interactions between the defects. 
In contrast, the highest energy configurations feature parallel vacancies that create a long-range structural instability.

Beyond three ligands, there is a combinatorial explosion and we become limited by our simulation cell size. 
However, we have considered some representative configurations. For acetic acid, removing four ligands equating to two BDC linkers per metal node, has no significant energy penalty (Figure \ref{figure3}f). This result agrees, at least qualitatively, with experiment, in that a large increase in surface area can be obtained by using acetic acid as a modulator to remove linkers from the structure. The removal of five and six linkers from the unit cell results in a small increase in defect energy per linker removal, before phonon stability, and therefore structure stability, is lost with the removal of 7--8 linkers for acetic acid compensation. For Cl$^{-}$/H$_{2}$O (Figure \ref{figure3}g) there is a reduction in energy per defect when removing seven and eight BDC linkers (\textit{i.e.} 3.5 -- 4 linkers per metal node), together with a phase change from cubic to monoclinic symmetry, which occurs in a similar manner to the breathing motion of ``winerack'' MOFs. There is also an increased structural flexibility, due to the high number of vacant ligand sites. The predicted phase change occurs at a very high concentration of defects and so may not be experimentally observable. Simulated powder X-ray spectra are given in the SI.\cite{fleming2005gdis} The symmetry reduction to monoclinic does not happen in the case of the acetic acid charge cap, since this is a bidentate ligand and the structural integrity of the cubic phase is maintained. 

A Boltzmann distribution for two and three linker vacancies shows that 99$\%$ of defects will be clustered at 300 K for the acetate and Cl$^{-}$/H$_{2}$O, respectively. 
Under equilibrium conditions, a distribution of isolated vacancies is unlikely and a dominant preference for clustered vacancy motifs would be expected, which is consistent with recent X-ray scattering analysis.\cite{cliffe2014correlated}

\subsection{Ordered defect structure: NU-1000}

A further simulation was performed for the OH$^{-}$/H$_{2}$O charge capping system with 8 linkers missing from the cubic unit cell. This corresponds to NU-1000, a MOF synthesised from a different Zr precursor. As an analysis of the energy required to form this structure, we repeat the removal of 1--8 linkers in the same manner as previously performed, but instead for the OH$^{-}$/H$_{2}$O charge capping. The final structure is equivalent to NU-1000 and was constructed along the highest symmetry path (the same path as was followed for the acetate and Cl/H$_{2}$O charge capping). Interestingly, we do not see the same phase change as was observed with the Cl/H$_{2}$O capping, instead hydrogen bonding between the hydroxyl groups and water maintain the cubic symmetry with only small structural distortions. The defect energy associated with the formation of NU-1000 (8 vacant linkers from the cubic unit cell) is similar to the cost of a single defect (see SI), highlighting the unusual tolerance of UiO-66 for high defect concentrations. We note that the defect energy for this charge capping considers the OH$^{-}$ capping source to be from the splitting of water. Synthesis methods for NU-1000 involve the use of a Zr-OH precursor, which offers an alternative OH$^{-}$ source. We therefore highlight the observed trend as being of interest rather than the specific energetics of ligand removal for making NU-1000.

\subsection{Molecular association in solution}
\begin{figure}
\centering
\includegraphics[width=8.6cm]{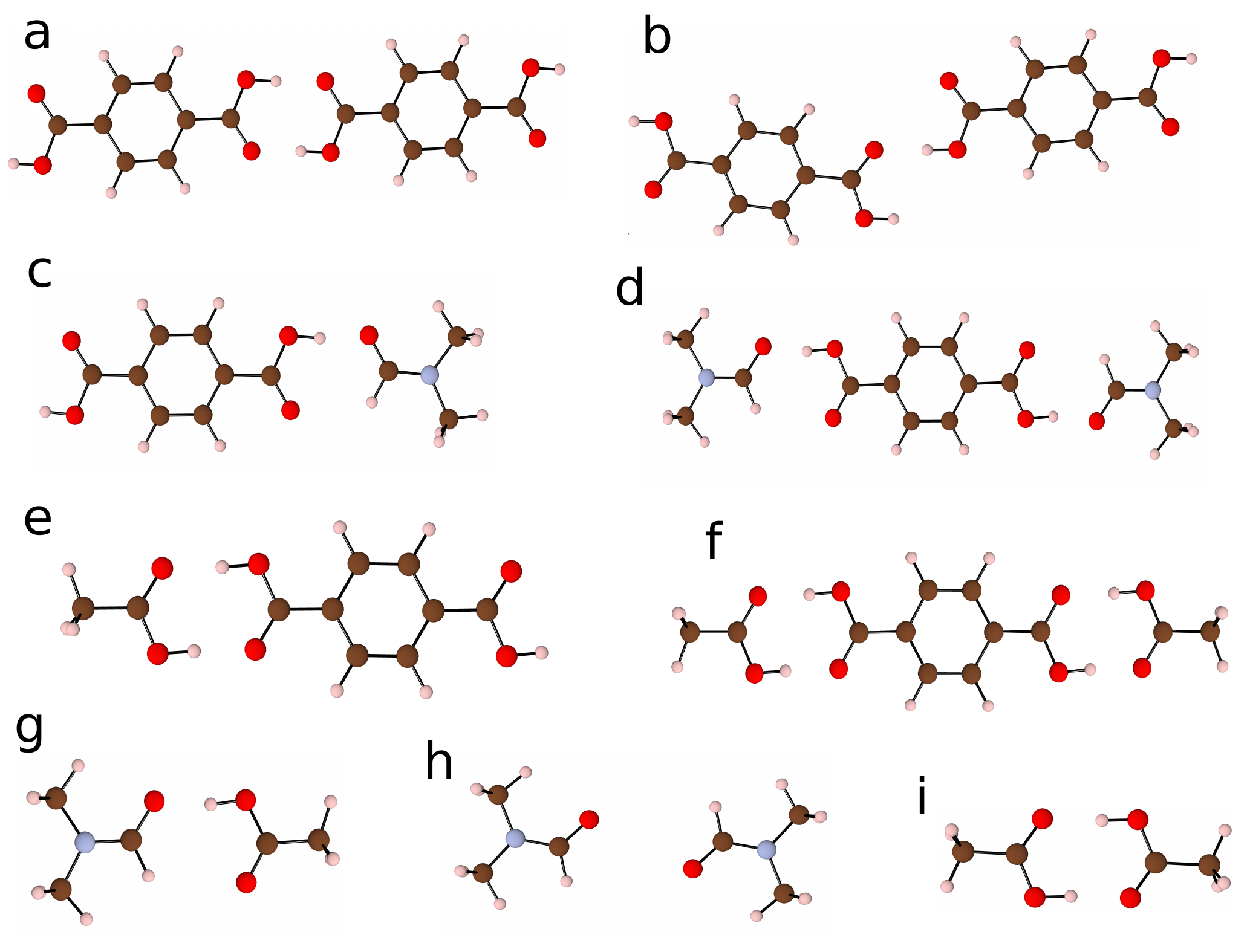}
\caption{Equilibrium geometries of molecular clusters for which binding energies are given in Table \ref{clustenergies}.}
\label{clusters}
\end{figure}

Due to the high concentration of defects predicted for UiO-66, we should consider processes beyond the typical dilute limit of non-interacting defects. Cluster formation following the removal and subsequent protonation of BDC may occur both in the framework, but also between the removed species in the solvent. Possible clusters that may form in solution are depicted in Figure \ref{clusters}. 
A strong binding energy of -104.7 kJmol$^{-1}$ between two acetic acid molecules and one BDC-H$_{2}$ linker has been calculated (Figure \ref{clusters}f).

Formation of molecular clusters in solution may provide an additional driving force for BDC linker to leave the UiO-66 framework when this acid is used as a modulator. 
Other clusters considered are shown to have a weaker binding energy between components (Table \ref{clustenergies}). 
Experimental evidence has been reported that even when synthesised without an acidic modulator UiO-66 can possess the missing linker defect at a low concentration. 
A contributing factor may be the strong calculated binding energy (-75.3 kJmol$^{-1}$) between DMF and BDC-H$_{2}$ (Figure \ref{clusters}d). 
The formation of this cluster can provide a thermodynamic driving force for a reduced number of linkers to be incorporated into the framework during the formation of UiO-66. 
The values reported are qualitative since hydrogen bonding between the solvent and molecule is not described in a continuum model.
An explicit solvent model could provide a more accurate description of aggregate formation in future studies.

\begin{table}[h]
\caption{Binding energies (after BSSE correction) of molecular clusters shown in Figure \ref{clusters} 
formed following linker removal from UiO-66 at 300 K (in DMF solvent).}
\centering
\begin{tabular}{c c c c c } 
\hline
Cluster & & & & $\Delta$E (kJmol$^{-1}$) \\
\hline
a & BDC & BDC & - & -47.6 \\
b & BDC & BDC & - & -22.5 \\
c & BDC &  DMF  & - & -29.5 \\
d & BDC & DMF & DMF & -75.3 \\
e & BDC & CH$_{3}$OOH & - & -52.8  \\
f & BDC & CH$_{3}$OOH &  CH$_{3}$OOH & -104.7   \\
g &  CH$_{3}$OOH & DMF & - & -38.3   \\
h &  DMF & DMF & - & -4.7   \\
i &  CH$_{3}$OOH & CH$_{3}$OOH & - & -56.1   \\
\hline
\end{tabular}
\label{clustenergies}
\end{table}

\subsection{Spectroscopic signatures}

The volume of the crystal lattice is found to increase and bulk modulus to decrease for the majority of capping models (see Table \ref{properties}). 
The single anion capping (Cl$^{-}$ and OH$^{-}$) is an exception as the anion effectively bridges between two metal centres, taking less physical space than BDC, and the lattice volume decreases.
The bulk moduli are all lower for the defect structures but remain within 5 GPa of pristine  UiO-66.
%

\begin{table}[h]
\caption{Structural and mechanical properties of pristine and defective UiO-66 with different capping mechanisms following the removal of a single BDC ligand.}
\centering
\begin{tabular}{c c c  } 
\hline
 Capping  &  Volume (\AA$^{3}$) & Bulk modulus (GPa)  \\
\hline
 UiO-66  			& 9120 & 23.04  \\
 Cl$^{-}$			& 9074 & 20.03  \\ 
 Cl$^{-}$/H$_{2}$O 	& 9126 & 20.77  \\
 Cl$^{-}$/DMF 		& 9132 & 21.15   \\
 OH$^{-}$ 			& 9092 & 20.19   \\
 OH$^{-}$/H$_{2}$O 	& 9137 & 20.67   \\
 OH$^{-}$/DMF 		& 9138 & 19.98  \\
  CH$_{3}$COO$^{-}$	& 9148 & 20.60  \\
\hline
\end{tabular}
\label{properties}
\end{table}

\begin{figure}
\centering
\includegraphics[width=8.3cm]{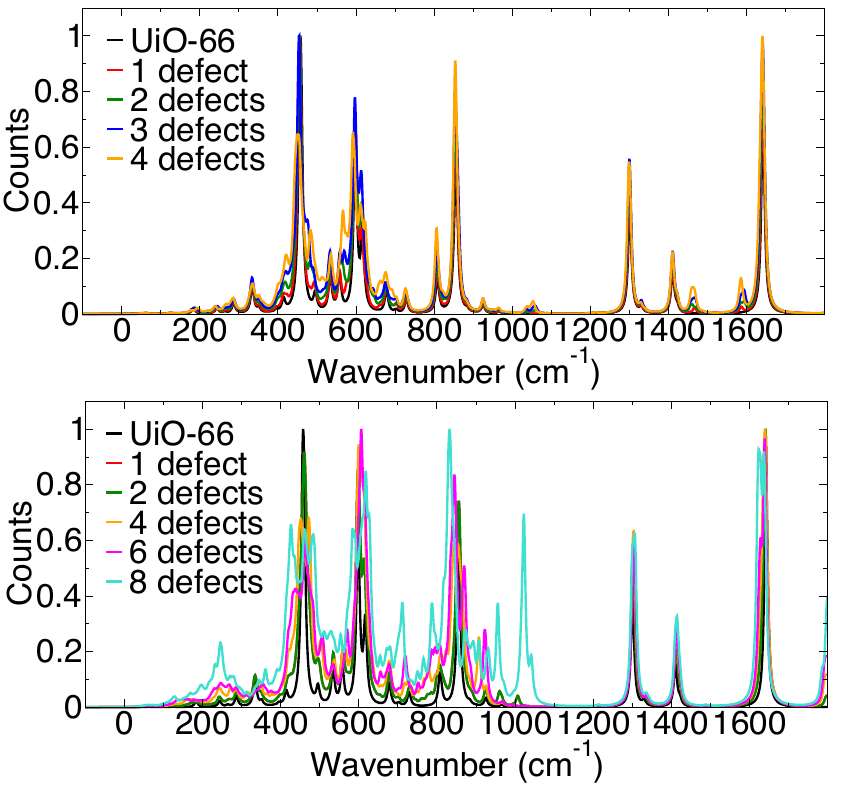}
\caption{Simulated IR spectra for pristine and defective UiO-66 with acetate (1--4 missing BDC linkers) (top) and Cl$^{-}$/H$_{2}$O (1,2,4,6 and 8 missing BDC linkers) (bottom) as the charge capping mechanism. IR spectra are plotted between 900--1800 cm$^{-1}$, which is the region where differences with defect concentrations are observed. A broadening factor of 10cm$^{-1}$ was applied.}
\label{ir1}
\end{figure}

A key question is whether the missing ligands have an observable spectroscopic signature. 
The simulated infrared (IR) spectra for 1--4 missing linkers for the two lowest energy charge capping mechanisms (acetate and Cl$^{-}$/H$_{2}$O) are presented in Figure \ref{ir1}. 
We highlight several important features for the identification of either charge cap. 
Firstly, for the acetate capping, acetate peaks are evident at 1463 cm$^{-1}$ and between 1583 -- 1586 cm$^{-1}$ due to the asymmetric and symmetric stretching of the C-O carboxylate bonds, respectively, which can be distinguished from the C-O carboxylate stretch of BDC, occurring between 1617-1650 cm$^{-1}$. 
The C-H bond stretch of acetate occurs at 2900 cm$^{-1}$, and the BDC C-H stretch at 2947 cm$^{-1}$. 
Additional peaks between 720 - 994 cm$^{-1}$ are associated with bending and twisting of the Zr node. Shoulder peaks are associated with the loss of symmetry at the Zr node, but are difficult to distinguish. For the Cl$^{-}$/H$_{2}$O charge cap, allocating specific frequencies is more difficult. As was the case for acetate, additional peaks between 500 - 900 cm$^{-1}$ are present due to the reduction in symmetry of the Zr node (as evident for 8 missing linkers in Figure \ref{ir1}). 
The Zr-Cl stretch is difficult to assign to one specific mode, but occurs in the same frequency range as the Zr-O stretches between 582 - 612 cm$^{-1}$. The most obvious difference for this system is the O-H bond stretch of water at 3378 cm$^{-1}$ (see SI for the full spectral range and comparisons to DFT calculations). 
The results suggest that high-resolution vibrational spectroscopy may provide the means to assign the local charge capping mechanism and give insights into defect concentrations. 
            
\section{Conclusion}

From an analysis of the defect chemistry of linker removal in UiO-66, we conclude that the lowest energy processes are for acetate and Cl$^{-}$/H$_{2}$O charge capping mechanisms. We show that H$_{2}$O capping at high concentrations results in an ordered-defect structure consistent with the NU-1000 framework.
A cluster between two acetic acid molecules and a protonated BDC linker is found to have a strong binding affinity and is a candidate product of ligand loss. 
The results are expected to be transferable to other UiO frameworks, with relevance to a wider range of hybrid organic-inorganic solids.

\acknowledgement
J.K.B is funded by the EPSRC (EP/G03768X/1). J.D.G thanks the Australian Research Council for funding under the Discovery Program, as well as the Pawsey Supercomputing Centre and NCI for the provision of computing resources. A.W. acknowledges support from the Royal Society. 
D.T. and K.L.S. are funded under ERC Starting Grant 277757 and J.M.S is funded under EPSRC grant no. EP/K004956/1. 
Access to the ARCHER supercomputer was facilitated through membership of the HPC Materials Chemistry Consortium (EP/L000202).

\bibliographystyle{rsc}
\bibliography{mofbib}

\suppinfo
Further methodological and computational details including a full breakdown of the defect free energies.



\end{document}